# PyR@TE 3

Lohan Sartore[a,*], Ingo Schienbein[a]

[a]*Laboratoire de Physique Subatomique et de Cosmologie, Université Grenoble-Alpes, CNRS/IN2P3, 53 Avenue des Martyrs, 38026 Grenoble, France*

**Abstract**

We present a new version of PyR@TE, a Python tool for the computation of renormalization group equations for general, non-supersymmetric gauge theories. Its new core relies on a recent paper by Poole & Thomsen [1] to compute the $\beta$-functions. In this framework, gauge kinetic mixing is naturally implemented, and the Weyl consistency relations between gauge, quartic and Yukawa couplings are automatically satisfied. One of the main new features is the possibility for the user to compute the gauge coupling $\beta$-functions up to the three-loop order. Large parts of the PyR@TE code have been rewritten and improved, including the group theory module PyLie. As a result, the overall performance in terms of computation speed was drastically improved and the model file is more flexible and user-friendly.

**NEW VERSION PROGRAM SUMMARY**

*Program Title:* PyR@TE 3
*Licensing provisions:* Apache 2.0
*Programming language:* Python 3
*Journal reference of previous version:* PyR@TE [1], PyR@TE 2 [2]
*Does the new version supersede the previous version?:* Yes.
*Reasons for new version:* The software was essentially rewritten and new functionalities were added. The performance in terms of computation speed was improved by a factor of 100 to 10000 compared to the previous version. The code now relies on Python 3 instead of the deprecated Python 2.
*Summary of revisions:* The core of the software was rewritten, based on a new formalism. One of the major new features is the possibility of computing the 3-loop RGEs for gauge couplings. The structure and the syntax of the model file were enhanced. The output of the software was improved.
*Nature of problem :* Computing the renormalization group equations for any renormalizable, 4-dimensional, non-supersymmetric quantum field theory.
*Solution method :* Group theory, tensor algebra

*Corresponding author.
*Email addresses:* sartore@lpsc.in2p3.fr (Lohan Sartore), schien@lpsc.in2p3.fr (Ingo Schienbein)

[2] F. Lyonnet, I. Schienbein, PyR@TE 2: A Python tool for computing RGEs at two-loop 213 181–196. doi:10.1016/j.cpc.2016.12.003.

## 1. Introduction

Renormalization group equations (RGEs) are an essential tool for modern high energy physics, as they provide the link between various energy scales in the framework of a given gauge theory. They allow, for instance, the study of vacuum stability, gauge coupling unification in the context of Grand Unified Theories (GUT), or any kind of high scale boundary conditions. Most generally, their importance is expected to keep growing in the field of high energy physics where probed scales are ever increasing compared to the electroweak (EW) scale. Thus, in the era of precision physics, the effect of the running couplings may tend to be systematically considered, either when it comes to perform precision tests of the Standard Model (SM) or to accurately constrain physics beyond the Standard Model (BSM).

RGEs for general gauge theories have been known for almost forty years at two-loop accuracy [2, 3, 4, 5, 6, 7, 8]. Since the derivation of the RGEs for specific models is often tedious and error prone in practice, these general expressions were implemented in the Mathematica package SARAH [9] and in the Python software PyR@TE [10, 11] allowing for an automatized computation of the $\beta$-functions. Over the years, the formalism was extended to take into account Abelian kinetic mixing [12, 13] and scalar mixing [14]. The application of the dummy field method, allowing for the determination of dimensionful couplings RGEs, was carried out rigorously and extensively reviewed in [14]. In addition, various mistakes and misprints were uncovered.

Recently, a new formalism for the RGEs of a general gauge theory was presented in [1]. In addition to systematically taking into account kinetic and scalar mixing, the full set of 3-loop $\beta$-functions for the gauge couplings was computed for the first time in the framework of a theory with a semi-simple gauge group (3-loop gauge coupling RGEs for theories based on a simple gauge group were first presented in [15]). The method used in this work relies on the local renormalization group [16, 17, 18] which reveals relations between the $\beta$-functions of gauge, Yukawa and scalar quartic couplings at different loop-orders: the so-called Weyl consistency conditions [17, 19, 20]. These relations constitute not only a powerful way of cross-checking existing results, but also provide a new tool for the determination of still unknown higher-loop order $\beta$-functions, as demonstrated in [21] where the 4-loop gauge $\beta$-functions of the Standard Model were computed for the first time. For these reasons we believe that this modern formalism might become the new paradigm in the field, thus motivating its implementation in the new version of PyR@TE. For this purpose, it is also necessary to know the RGEs for the dimensionful parameters in this formalism which can be found in [22].

In addition to the possibility of computing the gauge couplings RGEs up to the 3-loop order, inherent to the formalism described above, many improvements were made to the PyR@TE code. The latter was essentially rewritten, driven by the intention of improving its performances and overall ease of use. Several new features were also developed and will be extensively reviewed in the following.

The rest of this paper is organized as follows. Section 2 provides an introduction to the theoretical framework implemented in the new version of the software. Section 3 aims at serving as a self-contained user manual explaining how to download, install, and use PyR@TE 3. Next, in Sec. 4, we use various models for which we compare the $\beta$-functions obtained with

PyR@TE 3 with results in the literature or from PyR@TE 2. With very few exceptions, we find perfect agreement which validates our new tool which henceforth replaces the older versions of PyR@TE. Finally, in Sec. 5, we summarize the main new features of PyR@TE 3 and draw our conclusions. Some technical details have been relegated to the Appendix. In particular, we discuss the gauge kinetic mixing in more detail in Appendix A, the complete SM model file is listed in Appendix B, and a description of the group theory functionalities of PyR@TE 3 is provided in Appendix C.

## 2. Presentation of the formalism

We present in this section a brief overview of the adopted formalism as implemented in PyR@TE 3. The interested reader is invited to refer to [1] for any further details.

### 2.1. Definitions

Let us consider a general theory containing fermions and scalars transforming under various representations of a given semi-simple gauge group $\mathcal{G}$. Independently of the exact nature of the fermions (Dirac, Majorana or Weyl spinors) and scalars (complex or real) of the theory, we may always split them into their individual components and, without loss of generality, rearrange them in the form of two vectors $\psi_i$ and $\phi_a$, of Weyl fermions and real scalars, respectively. The fermionic index $i$ ranges from 1 to $n_f$ and the scalar index $a$ from 1 to $n_s$. In this context, the non-kinetic part of the Lagrangian density may be written as

$$\mathcal{L} = -\frac{1}{2}\left(Y_{aij}\psi_i\psi_j\phi_a + \text{h.c.}\right) - \frac{1}{2}\left(M_{ij}\psi_i\psi_j + \text{h.c.}\right) \\ -\frac{1}{2}\mu_{ab}\phi_a\phi_b - \frac{1}{3!}t_{abc}\phi_a\phi_b\phi_c - \frac{1}{4!}\lambda_{abcd}\phi_a\phi_b\phi_c\phi_d \,. \quad (1)$$

Closely following [1], we further generalize the notation defining the Majorana spinor

$$\Psi_i \equiv \begin{pmatrix} \psi \\ \psi^\dagger \end{pmatrix}_i, \quad (2)$$

and the associated Yukawa and fermion mass couplings tensors

$$y_{aij} \equiv \begin{pmatrix} Y_a & 0 \\ 0 & Y_a^* \end{pmatrix}_{ij}, \qquad m_{ij} \equiv \begin{pmatrix} M & 0 \\ 0 & M^* \end{pmatrix}_{ij}. \quad (3)$$

In the two definitions above and from now on, the fermionic indices range from 1 to $2n_f$ since they run on both fermions and antifermions of the model. The Lagrangian density (1) now takes the form

$$\mathcal{L} = -\frac{1}{2}y_{aij}\Psi_i\Psi_j\phi_a - \frac{1}{2}m_{ij}\Psi_i\Psi_j \\ -\frac{1}{2}\mu_{ab}\phi_a\phi_b - \frac{1}{3!}t_{abc}\phi_a\phi_b\phi_c - \frac{1}{4!}\lambda_{abcd}\phi_a\phi_b\phi_c\phi_d \,. \quad (4)$$

Now turning to the gauge sector of the theory, we introduce the compact and very convenient notation developed in [1], starting from the decomposition of the semi-simple gauge group in its $N$ individual gauge factors:

$$\mathcal{G} = \prod_{u=1}^{N} \mathcal{G}_u \,, \quad (5)$$

where the $\mathcal{G}_u$'s are compact Lie groups with dimension $d_u$. Absorbing the coupling constants $g_u$ in the definition of the gauge fields (the procedure in the case of Abelian kinetic mixing is detailed in Appendix A), the covariant derivatives of the Weyl fermions $\psi_i$ and real scalars $\phi_a$ are defined as

$$D_\mu \psi_i = \partial_\mu \psi_i - i \sum_{u=1}^{N} \sum_{A_u=1}^{d_u} V_\mu^{A_u} \left(T_\psi^{A_u}\right)_{ij} \psi_j, \tag{6}$$

$$D_\mu \phi_a = \partial_\mu \phi_a - i \sum_{u=1}^{N} \sum_{A_u=1}^{d_u} V_\mu^{A_u} \left(T_\phi^{A_u}\right)_{ab} \phi_b. \tag{7}$$

For fermions, we make use of the definition (2) to define the gauge generators in the new $(2n_f)$-dimensional fermionic space:

$$\left(T^{A_u}\right)_{ij} \equiv \begin{pmatrix} T_\psi^{A_u} & 0 \\ 0 & -\left(T_\psi^{A_u}\right)^* \end{pmatrix}_{ij}, \tag{8}$$

such that the covariant derivative of $\Psi_i$ reads

$$D_\mu \Psi_i = \partial_\mu \Psi_i - i \sum_{u=1}^{N} \sum_{A_u=1}^{d_u} V_\mu^{A_u} \left(T^{A_u}\right)_{ij} \Psi_j. \tag{9}$$

Another object that we need to define, and which we will make an extensive use of, is the gauge-coupling matrix $G_{AB}^2$. In a theory without kinetic mixing in the Abelian sector (*i.e.* with at most one $U(1)$ gauge factor), the latter is simply defined as the diagonal matrix

$$G_{AB}^2 = g_u^2 \delta^{A_u B_u}, \tag{10}$$

where we used the convention [1]

$$\sum_A = \sum_u \sum_{A_u}^{d_u}. \tag{11}$$

Now considering the most general theory with $p$ Abelian gauge factors, the matrix $G^2$ develops off-diagonal components and has the following form

$$G_{AB}^2 = \begin{pmatrix} \begin{matrix} h_{11} & \cdots & h_{1p} \\ \vdots & \ddots & \vdots \\ h_{p1} & \cdots & h_{pp} \end{matrix} & & & \\ & g_{p+1}^2 & & \\ & & \ddots & \\ & & & g_n^2 \end{pmatrix}_{AB} \equiv \begin{pmatrix} H^2 & 0 \\ 0 & G_{\text{na}}^2 \end{pmatrix}_{AB}. \tag{12}$$

In the above parametrization, $G_{\text{na}}^2$ is simply a diagonal matrix populated by the non-Abelian gauge couplings (see eq. (10)) and $H^2 = G_{\text{mix}} G_{\text{mix}}^{\text{T}}$. As explained in Appendix A, without loss of generality, we will in practice express $G_{\text{mix}}$ as an upper-triangular $p \times p$ matrix. For instance, if the Abelian sector of the theory is $U(1)_Y \times U(1)_{B-L}$, PyR@TE will internally construct the matrix $G_{\text{mix}}$ such that

$$G_{\text{mix}} = \begin{pmatrix} g_{U1Y} & g_{12} \\ 0 & g_{U1BL} \end{pmatrix}, \tag{13}$$

and compute the $\beta$-functions of the $p(p+1)/2 = 3$ gauge couplings defined thereby. We emphasize that, independently of the presence of Abelian mixing in the gauge sector, the matrix $G_{AB}^2$ is always symmetric and positive definite.

To conclude this section, we eventually present the full form of the most general Lagrangian density [1]:

$$\begin{aligned}\mathcal{L} = &-\frac{1}{4}\left(G^{-2}\right)_{AB} F^A_{\mu\nu} F^{B\mu\nu} + \frac{1}{2}(D_\mu\phi)_a(D^\mu\phi)_a + \frac{i}{2}\Psi^T \begin{pmatrix} 0 & \sigma^\mu \\ \bar{\sigma}^\mu & 0 \end{pmatrix} D_\mu \Psi \\ &-\frac{1}{2}y_{aij}\Psi_i\Psi_j\phi_a - \frac{1}{2}m_{ij}\Psi_i\Psi_j \\ &-\frac{1}{2}\mu_{ab}\phi_a\phi_b - \frac{1}{3!}t_{abc}\phi_a\phi_b\phi_c - \frac{1}{4!}\lambda_{abcd}\phi_a\phi_b\phi_c\phi_d.\end{aligned} \tag{14}$$

### 2.2. $\beta$-functions

The expressions of the $\beta$-functions for the dimensionless (marginal) couplings are given by [1]:

$$\beta_{AB} \equiv \frac{dG_{AB}^2}{dt} = \frac{1}{2}\sum_{\text{perm}}\sum_\ell \frac{1}{(4\pi)^{2\ell}} G_{AC}^2 \beta_{CD}^{(\ell)} G_{DB}^2, \tag{15}$$

$$\beta_{aij} \equiv \frac{dy_{aij}}{dt} = \frac{1}{2}\sum_{\text{perm}}\sum_\ell \frac{1}{(4\pi)^{2\ell}} \beta_{aij}^{(\ell)}, \tag{16}$$

$$\beta_{abcd} \equiv \frac{d\lambda_{abcd}}{dt} = \frac{1}{4!}\sum_{\text{perm}}\sum_\ell \frac{1}{(4\pi)^{2\ell}} \beta_{abcd}^{(\ell)}, \tag{17}$$

where $t = \ln \mu/\mu_0$, $\mu$ standing for the renormalization scale and $\mu_0$ some fixed, arbitrary energy scale. The permutations are respectively understood to be taken among the two gauge indices, the two fermionic indices of the Yukawa couplings and the four scalar indices of the quartic couplings. As usual, the $\beta$-functions are given in the form of a perturbative expansion with $\ell$ denoting the loop-order. Based on a diagrammatic approach, the $\beta^{(\ell)}$ are expressed as a sum of individual contributions, each weighted by a coefficient (see Eq. (5.2) in [1]) whose actual value depends on the renormalization scheme. For instance, at the 1-loop order, $\beta_{AB}^{(1)}$ is expressed as

$$\beta_{AB}^{(1)} = \mathfrak{g}_1^{(1)}\left[C_2(G)\right]_{AB} + \mathfrak{g}_2^{(1)}\left[S_2(F)\right]_{AB} + \mathfrak{g}_3^{(1)}\left[S_2(S)\right]_{AB}, \tag{18}$$

where the numerical coefficients are given in the $\overline{\text{MS}}$ scheme by

$$\mathfrak{g}_1^{(1)} = -\frac{22}{3}, \qquad \mathfrak{g}_2^{(1)} = \frac{2}{3}, \qquad \mathfrak{g}_3^{(1)} = \frac{1}{3}, \tag{19}$$

and where the three tensor structures encode the model-dependent contributions from the gauge boson, fermion and scalar loops, respectively. Their definitions are given in Appendix A of [1], which we reproduce here for completeness:

$$[C_2(G)]_{AB} = f^{ACD}f^{CDB}, \quad [S_2(F)]_{AB} = \text{Tr}\left[T^A T^B\right], \quad [S_2(S)]_{AB} = \text{Tr}\left[T_\phi^A T_\phi^B\right]. \tag{20}$$

Evaluating Eq. (18) in a model based on a simple gauge group with coupling $g$ yields the well-known expression (valid for two-component fermions)

$$\frac{dg^2}{dt} = \frac{g^4}{(4\pi)^2}\left(-\frac{22}{3}C_2(G) + \frac{4}{3}S_2(F) + \frac{1}{3}S_2(S)\right). \quad (21)$$

The full set of gauge, Yukawa and quartic RGEs in this formalism is presented, up to the respective loop orders 3-2-2, in Appendix A of [1] along with the list of numerical coefficients given in the $\overline{\text{MS}}$ scheme. These expressions were directly implemented in PyR@TE 3 and constitute the core of the algorithm used to produce the (dimensionless) $\beta$-functions of a given model.

*2.3. Dimensionful couplings*

Thanks to a proper application of the so-called dummy field method [14], one may obtain the RGEs of the dimensionful couplings (fermion mass, scalar mass and scalar trilinear couplings) of a theory based on the known expressions for the dimensionless ones. The formalism presented in the previous sections is particularly well-suited for such a procedure, which relies on a diagrammatic approach as well. For instance, one can derive the contribution to a fermion mass $\beta$-function by amputating the external scalar leg of a $\ell$-loop Yukawa coupling diagram involving a dummy scalar field introduced at the Lagrangian level. We will not go through this procedure in detail since it is extensively documented in [14] but simply mention that we used it to derive the full set of dimensionful RGEs of a general gauge theory up to the 2-loop order, based on the results of [1]. Following Eqs. (15) - (17), the expression of the $\beta$-functions for fermion mass, trilinear and scalar mass couplings are respectively expressed as:

$$\beta_{ij} \equiv \frac{dm_{ij}}{dt} = \frac{1}{2}\sum_{\text{perm}}\sum_{\ell}\frac{1}{(4\pi)^{2\ell}}\beta_{ij}^{(\ell)}, \quad (22)$$

$$\beta_{abc} \equiv \frac{dt_{abc}}{dt} = \frac{1}{3!}\sum_{\text{perm}}\sum_{\ell}\frac{1}{(4\pi)^{2\ell}}\beta_{abc}^{(\ell)}, \quad (23)$$

$$\beta_{ab} \equiv \frac{d\mu_{ab}}{dt} = \frac{1}{2}\sum_{\text{perm}}\sum_{\ell}\frac{1}{(4\pi)^{2\ell}}\beta_{ab}^{(\ell)}. \quad (24)$$

The complete set of expressions of the $\beta^{(\ell)}$ for dimensionful couplings is given in Ref. [22].

## 3. PyR@TE 3 : an overview

*3.1. Download and installation*

The new version of PyR@TE may be downloaded from the following GitHub web page:

```
https://github.com/LSartore/pyrate
```

No further installation procedure is needed and the user may start working in PyR@TE's main folder. However, PyR@TE relies on a number of Python modules which need to be installed by the user if not already present on his/her machine. We show here the full list of dependencies:

- `Python` $\geq$ 3.6
- `PyYAML` $\geq^*$ 5.3
- `Sympy` $\geq$ 1.5
- `h5py` $\geq^*$ 2.10
- `Numpy` $\geq^*$ 1.18
- `Scipy` $\geq^*$ 1.4
- `Matplotlib` $\geq^*$ 3.1

Version requirements marked with an asterisk are not *critical* requirements, in the sense that PyR@TE is likely to work properly with older versions of these modules. However, we note that during its development, PyR@TE was only tested with the above listed versions. We also note that `Scipy` and `Matplotlib` are solely required to run the Python output of PyR@TE (see section 3.4).

*3.2. Definition of the model*

The information on the particle physics model is contained in the so-called model file, which is the starting point of any computation performed by PyR@TE. The overall structure of the model file is essentially based on the previous versions [10, 11] of the software. However, a number of improvements were made that we will review in the following. Let us begin this discussion with, as usual, the example of the Standard Model. The full SM model file is provided in Appendix B. In the following we will go through this model file step by step, making comments wherever useful.

**General information** – Three fields can be provided in this section, namely the `Author`, `Date` of creation and `Name` of the model file. The first two are essentially informative while the last one will be used by PyR@TE to generate its output (see section 3.4).

```
Author: Lohan Sartore
Date: 08.06.2020
Name: SM
```

**Gauge groups** – This is the first essential information to provide to PyR@TE. Here the full gauge group of the SM is $U(1)_Y \times SU(2)_L \times SU(3)_c$:

```
Groups: {U1Y: U1, SU2L: SU2, SU3c: SU3}
```

Each gauge factor must be given a custom label, and is described by the usual name of the associated Lie group. Possible choices cover the entirety of the gauge groups associated with a simple Lie algebra, namely $U(1)$, $SU(N)$, $SO(N)$, $Sp(N)$ and the five exceptional groups $G_2$, $F_4$ and $E_{6,7,8}$. We emphasize that this is an improvement compared to PyR@TE 2, where symplectic and exceptional groups were not implemented.

**Field content** – The next step consists in defining the particle content of the model. Three kinds of fields may be implemented in the model file, namely `Fermions`, `RealScalars` and `ComplexScalars`:

```
Fermions: {
    Q  : {Gen: 3, Qnb: {U1Y: 1/6, SU2L: 2, SU3c: 3}},
    L  : {Gen: 3, Qnb: {U1Y: -1/2, SU2L: 2, SU3c: 1}},
    uR : {Gen: 3, Qnb: {U1Y: 2/3, SU2L: 1, SU3c: 3}},
    dR : {Gen: 3, Qnb: {U1Y: -1/3, SU2L: 1, SU3c: 3}},
    eR : {Gen: 3, Qnb: {U1Y: -1, SU2L: 1, SU3c: 1}},
}

RealScalars: {
}

ComplexScalars: {
    H : {RealFields: [Pi, Sigma], Norm: 1/sqrt(2), Qnb: {U1Y: 1/2, SU2L: 2, SU3c: 1}},
}
```

In every case the user must specify the quantum numbers of each particle under the gauge group of the model. For Abelian gauge factors, we call *quantum number* the charge of the fields (*e.g.* the hypercharge under $U(1)_Y$). For non-Abelian gauge factors, it corresponds instead to the irreducible representation under which the field transforms. Most of the time, the usual notation based on the dimension of the representation (*e.g.* **2** of $SU(2)$, **3** of $SU(3)$, ...) is sufficient to identify unambiguously a given representation. In this notation, the conjugate representations are indicated by a negative quantum number in the model file. For instance, it is understood that `Qnb: {..., SU3c : -3}` refers to the anti-fundamental representation of $SU(3)_c$. In addition, the quantum number `1` indicates that the field transforms in the trivial representation, *i.e.*, is unaffected by the gauge transformations of the associated gauge factor. Another way to indicate that a field is not charged under one or several gauge factors is to simply omit them in the definition of the quantum numbers. For instance, it is understood that `eR` is a singlet under both `SU2L` and `SU3c` if the user writes:

```
    eR : {Gen: 3, Qnb: {U1Y : -1}}
```

This remark also applies to Abelian gauge factors, in which case the charge of the corresponding field is automatically set to `0`.

In a situation where a given gauge group possesses several representations with the same dimension (*e.g.* the **15** and **15′** of $SU(3)$), the usual notation breaks down and the user must use the Dynkin labels notation instead. To keep this presentation straightforward we defer the discussion on Dynkin labels and their implementation in the model file to Appendix C.

In addition to the quantum numbers, the `Fermions` of the model may be assigned a generation number. This number can either be a positive integer or a symbolic number, for instance: `Q : {Gen: nG, ...}`. In the latter case the $\beta$-functions will be expressed in terms of the symbolic generation numbers they may explicitly depend on.

For `ComplexScalars`, the user must indicate the decomposition of the fields in terms of a real and imaginary part. To this end, he/she must use the `RealFields` and `Norm` keywords.

In our case, the complex Higgs doublet is expressed as

$$H = \frac{1}{\sqrt{2}}\left(\Pi + i\,\Sigma\right). \tag{25}$$

We note that there is no need to define the real fields beforehand in the `RealScalars` section of the model file. We also want to draw the reader's attention to the influence of the norm of complex scalars on the actual expression of the resulting $\beta$-functions: varying the norm of a complex scalar field can be seen as a rescaling of its real components, or, equivalently, a rescaling of the couplings of the model.

Finally, since the SM contains no real scalars (other than the real components of the Higgs doublet), we show here an example of syntax:

```
RealScalars: {
    S : {Qnb: {U1Y: 1/2, SU2L: 1, SU3c: 1}}
}
```

A short-hand syntax may also be employed, removing the `Qnb` keyword:

```
RealScalars: {
    S : {U1Y: 1/2, SU2L: 1, SU3c: 1}
}
```

A last important point concerns the conjugate fields of the model. In contrast to the previous version of PyR@TE, the anti-particles should never be defined in the model file. Instead, every fermion and complex scalar is automatically assigned a conjugate counterpart which can be accessed appending `bar` to its name. For instance, in our case, `Qbar` and `Hbar` would respectively correspond to the fields $\bar{Q}$ and $H^\dagger$ of the Standard Model.

**Potential** – This section contains the expression of the Lagrangian density of the theory. Since kinetic terms are not needed in PyR@TE, the only types of couplings which must be defined are:

- The Yukawa couplings (`Yukawas`)

- The quartic scalar couplings (`QuarticTerms`)

- The trilinear scalar couplings (`TrilinearTerms`)

- The scalar mass couplings (`ScalarMasses`)

- The fermion mass couplings (`FermionMasses`)

The syntax used in these five sections to define the Lagrangian was revisited in PyR@TE 3. In order for the user to have a full control over the expressions of the various terms, the new syntax consists in writing the terms as an explicit mathematical expression with contracted indices. As an illustration, the down-type Yukawa coupling of the Standard Model,

$$Y_{d\,f_1,f_2}\,\bar{Q}_{f_1,i,a}\,H^i\,d^a_{R,f_2} + \text{h.c.}, \tag{26}$$

is implemented in the model file as:

```
Yukawas: {
    Yd : Qbar[i,a] H[i] dR[a]
}
```

A few important remarks are in order.

First, in the definition of the Yukawa couplings (and fermion masses), the flavor indices are implicit. Consequently, the order in which the two fermions appear in the expression has an influence on the structure of the Yukawa matrices. In the example above, it is understood that `Qbar` is contracted with the first flavor index of `Yd` and `dR` with the second one.

Then, a field which transforms non trivially under more than one non-Abelian gauge factor will carry more than one index. In this case, the order in which the indices should be written is based on the order in which the gauge factors were defined. Therefore, in the above example, the `SU2` index of `Qbar` comes before the `SU3` index.

Finally, we note that the hermitian conjugate is automatically inferred by PyR@TE and therefore should not need to be defined explicitly. This behavior always applies to Yukawa and fermion mass couplings, but also to couplings involving scalar fields. In the latter case however, the user has the choice to include or not the conjugate couplings in the model file. Let us consider as an illustration the $\lambda_5$ quartic coupling of the Two-Higgs-doublet model:

$$\frac{1}{2}\left[\lambda_5 \left(\phi_1^\dagger \phi_2\right)^2 + \lambda_5^* \left(\phi_2^\dagger \phi_1\right)^2\right]. \tag{27}$$

There are essentially two ways of implementing this term:

```
QuarticCouplings: {
    # First possibility - the conjugate counterpart is explicitly defined
    lambda5 : 1/2 (phi1bar[i] phi2[i])**2,
    lambda5star : 1/2 (phi2bar[i] phi1[i])**2

    # Second possibility - let PyR@TE infer the conjugate automatically
    lambda5 : 1/2 (phi1bar[i] phi2[i])**2
}
```

In the first implementation, the $\lambda_5^*$ term is explicitly defined. In this situation, the suffix `star` must be appended to the name of the coupling (here, `lambda5star`) to enforce the relation of conjugation among the two couplings. In the second implementation, where the conjugate counterpart is omitted, PyR@TE will automatically detect that the term is not invariant under complex conjugation, and consequently generate the $\lambda_5^*$ term when constructing the Lagrangian of the model.

Finally, there may be situations where $\lambda_5$ is assumed real. In this case, the natural way to proceed is to set $\lambda_5^* = \lambda_5$ in Eq. (27), leading to the first implementation shown below. Equivalently, the assumption[1] `real` may be used to achieve the same outcome:

---

[1] More detail about assumptions is given below.

```
QuarticCouplings: {
    # lambda5 is assumed real
    lambda5 : 1/2 (phi1bar[i] phi2[i])**2 + 1/2 (phi2bar[i] phi1[i])**2

    # lambda5 is assumed real, shorthand
    lambda5 : {1/2 (phi1bar[i] phi2[i])**2 , real}
}
```

**Definitions** – In order to help the user deal with this new syntax and produce a clear and readable model file, a new section was introduced in PyR@TE 3. This is the `Definitions` section, which must be implemented inside the `Potential` part of the model file (see the full SM model file in Appendix B). In this section, the user may define quantities that will be used in the expression of the Lagrangian density. For instance, in the SM, one has to introduce the conjugated Higgs field:

$$\tilde{H} \equiv \varepsilon H^\dagger \,, \tag{28}$$

where $\varepsilon = i\sigma_2$ is the Levi-Civita tensor of rank 2. Such an auxiliary quantity may be defined in the model file the following way:

```
Definitions: {
    Htilde[i] : Eps[i,j]*Hbar[j]
}
```

The tensor `Eps` is a pre-defined object in PyR@TE 3. Levi-Civita tensors of rank higher than 2 may be defined in the same fashion, increasing the number of indices accordingly. We emphasize that the gauge indices must always appear explicitly, both on the right- and left-hand side. According to the usual convention, repeated indices will be summed over internally. Any object defined this way can now be used in the Lagrangian expression. In our case, the up-type Yukawa couplings may be written as:

```
Yukawas: {
    Yu : Qbar[i,a] Htilde[i] uR[a],
# ... is equivalent to ...
    Yu : Qbar[i,a] Eps[i,j] Hbar[j] uR[a],
}
```

Not only the `Definitions` sections may help produce a clear and structured model file, but it also introduces two additional features, namely:

1. The use of gauge group generators in the expression of the Lagrangian,

2. The possibility of using Clebsch-Gordan coefficients (CGCs) to produce terms that cannot be expressed simply in terms of products of fields with contracted indices.

Both rely on the group-theoretical module PyLie, first introduced in PyR@TE 2 [11]. This module underwent a number of modifications and improvements presented in Appendix C. We also defer to this appendix the treatment of the CGCs and their implementation in the model file. Here we focus on the first feature, namely the definition of gauge generators in the model file.

As an illustration, let us consider a toy model in which the scalar sector of the SM is extended by a complex $SU(2)$ triplet $\delta$. It is convenient to rewrite this triplet in the form of a $2 \times 2$ matrix by contraction with the generators of the fundamental representation:

$$\Delta \equiv t_a \, \delta^a. \tag{29}$$

In the following, we demonstrate how to use the `Definitions` section to implement the quartic part of the scalar potential:

$$V = \lambda_1 \left(H^\dagger H\right)^2 + \lambda_2 \text{Tr}(\Delta^\dagger \Delta) H^\dagger H + \lambda_3 H^\dagger \Delta \Delta^\dagger H + \lambda_4 \text{Tr}(\Delta^\dagger \Delta)^2 + \lambda_5 \text{Tr}(\Delta^\dagger \Delta \Delta^\dagger \Delta). \tag{30}$$

First, we have to introduce the generators $t_a$ in order to define the matrix $\Delta$. The general syntax to define the generators of a given representation is `t(group, representation)`, where the representation may be labeled either by its dimension or its Dynkin labels. We note that the generators thus defined carry three indices with respective ranges $D$, $N_r$ and $N_r$, where $D$ is the dimension of the Lie group (*i.e.* of its adjoint representation) and $N_r$ that of the considered representation. The matrix $\Delta$ and its hermitian conjugate $\Delta^\dagger$ may therefore be implemented as[2]:

```
Definitions: {
    # Define the generators of the fundamental rep of SU2 (indices are implicit)
    tFund : t(SU2, 2),

    # Define the matrix Delta and its adjoint
    Delta[i,j] : tFund[a,i,j] delta[a],
    DeltaDag[i,j] : tFund[a,i,j] deltabar[a]

    # We may also define the traces
    Tr2 : DeltaDag[i,j] Delta[j,i],
    Tr4 : DeltaDag[i,j] Delta[j,k] DeltaDag[k,l] Delta[l,i]
}
```

Having also pre-defined the traces $\text{Tr}(\Delta^\dagger \Delta)$ and $\text{Tr}(\Delta^\dagger \Delta \Delta^\dagger \Delta)$, the scalar potential (30) can now be implemented in a simple, concise form:

```
QuarticTerms: {
    lambda1 : (Hbar[i] H[i])**2,
    lambda2 : Hbar[i] H[i] Tr2,
    lambda3 : Hbar[i] Delta[i,j] DeltaDag[j,k] H[k],
    lambda4 : Tr2**2,
    lambda5 : Tr4
}
```

**Assumptions** – In some cases, the user may want to assume some particular properties for the Yukawa (or fermion mass) matrices. Four different assumptions can be implemented for such couplings in the model file, namely: `real`, `symmetric`, `hermitian` and `unitary`. To impose one or more of these properties to a Yukawa matrix, the general syntax is:

---

[2] In this example, we used the fact that $SU(N)$ generators are hermitian. In general, we may conjugate the representation matrices appending `bar` to the name of the generators. For instance, `tFundbar[a,j,i]` would correspond to $(t_a^*)_i^j = (t_a^\dagger)_j^i$.

```
coupling: {expression, assumption1, assumption2, ...}
```

For instance, let us assume a real and symmetric Yukawa matrix in the leptonic sector of the SM:

```
Yukawas: {
    # Without assumptions
    Ye : Lbar[i] H[i] eR

    # With assumptions
    Ye : {Lbar[i] H[i] eR, real, symmetric}
},
```

Based on these assumptions, PyR@TE will automatically perform the appropriate simplifications in the resulting RGEs. In the case illustrated above, $Y_e^\dagger$ would be systematically simplified as $Y_e$. We note that in some cases, such assumptions should be necessarily included in order to guarantee explicit gauge invariance of the Lagrangian, checked by PyR@TE 3 using the appropriate options (see section 3.3).

As stated previously, the `real` assumption may also be used for quartic, trilinear and scalar mass couplings.

At this point, it is important to note that depending on the model considered, such assumptions might not be preserved along the RG-flow. This may happen for instance if a coupling which is assumed real has a $\beta$-function with a non vanishing imaginary part. In such cases, the simplifications made in the expressions of the $\beta$-functions would not be valid at all scales and would generate an inconsistent RG-flow.

Finally, another possible assumption, `squared`, concerns only the scalar mass terms, and allows to make the distinction between the two following notations:

$$\mathcal{L} \supset \mu \phi^\dagger \phi \quad \text{and} \quad \mathcal{L} \supset \mu^2 \phi^\dagger \phi. \tag{31}$$

The former case, where the mass dimension of $\mu$ equals 2, is the one assumed in PyR@TE by default. In the latter case, the `squared` keyword may be added in order to indicate that the scalar mass coupling has a mass dimension of 1:

```
ScalarMasses: {
    mu : {Phibar[i] Phi[i], squared}
}
```

In this case, PyR@TE will internally use the relation

$$\beta(\mu) = \frac{1}{2\mu}\beta(\mu^2) \tag{32}$$

to infer the value of $\beta(\mu)$.

**Vacuum-expectation values** – The 2-loop RGEs of vacuum-expectation values (VeVs) were implemented in PyR@TE 3, based on the $\overline{\text{MS}}$ expressions of Refs. [23, 24]. To define a VeV, the only thing to do is to identify the real scalar component which develops a non-zero expectation value. For instance, considering that in the SM the VeV is developed by the real part of the second component of the Higgs doublet, we simply have to write:

```
Vevs: {
    v : Pi[2]
}
```

Since VeVs break gauge invariance, the resulting RGEs are given in a general $R_\xi$ gauge and therefore explicitly depend on the $\xi$ parameter. If the user wishes to fix the gauge, he/she may use the `GaugeParameter` keyword:

```
# Let's work in the Landau gauge
GaugeParameter: 0
```

**Anomalous dimensions** – The 2-point anomalous dimensions of scalars and fermions may be computed by PyR@TE. This is achieved by adding the optional `ScalarAnomalous` and `FermionAnomalous` sections in the model file. In order to compute the anomalous dimensions of one or several pairs of fields, the following syntax may be used:

```
ScalarAnomalous: {
    (Pi[1], Pi[1]),
    (Sigma[2], Sigma[2])
}

FermionAnomalous: {
    (Q[1,1], Q[1,1]),
    (eR, eR),
    (uR[1], dR[1])
}
```

We emphasize that the fields used as an input for computing the anomalous dimensions are the individual components of the gauge eigenstates, and not the fields themselves (except of course if a given field is a gauge singlet).

To compute all the possible anomalous dimensions of the model, the user may instead use the keyword `all` :

```
ScalarAnomalous : all
FermionAnomalous : all
```

In this case, PyR@TE will display in its output all the non-vanishing anomalous dimensions of the fields of the model.

We note finally that the anomalous dimensions depend in general on the gauge fixing parameter $\xi$. If the gauge was fixed using the `GaugeParameter` keyword introduced above, its value will be substituted accordingly in the expression of the anomalous dimensions.

**Substitutions** – Another new feature in PyR@TE 3 is the possibility of performing various kinds of substitutions after the computation of the RGEs is carried out. To this end, the `Substitutions` section can be implemented in the model file. The possible types of substitutions which may be performed are listed and exemplified below.

(1) The first type of substitution consists in renaming some of the couplings. This is mainly useful to rename the gauge couplings whose names are set by default to `g[GroupName]`. For instance, in the SM, the three gauge couplings would respectively be labeled as `gU1Y`, `gSU2L`, `gSU3c`. If one wishes to rename them to `g1`, `g2` and `g3`, the following syntax must be used:

```
Substitutions: {
    gU1Y : g1,
    gSU2L : g2,
    gSU3c : g3
}
```

(2) Another possibility is to apply GUT normalization factors to some of the gauge couplings. The following code indicates that the normalization $g_1 \to \sqrt{5/3}\, g_1$ must be adopted and substituted in the $\beta$-functions:

```
Substitutions: {
    g1 : sqrt(5/3)*g1
}
```

(3) In some cases, the user may want to constrain the form of the Yukawa (or fermion mass) matrices, for instance neglecting the first two generations or the off-diagonal couplings. The examples below show how to proceed in these two cases:

```
Substitutions: {
  # First example - Neglect the first and second generations
    Yu : [[0, 0,  0  ],
          [0, 0,  0  ],
          [0, 0, 'yt']],
    # A short-hand notation may be employed for diagonal matrices
    Yd : [0, 0, 'yb'],
    Ye : [0, 0, 'ytau'],

  # Second example - Neglect the off-diagonal couplings
    Yu : ['yu', 'yc', 'yt'],
    Yd : ['yd', 'ys', 'yb'],
    Ye : ['ye', 'ymu', 'ytau'],
}
```

It is important to note that the newly defined couplings (*e.g.* `'yt'`, `'yb'`, ...) are put inside quotation marks `''`. Furthermore, the user must be aware that in some cases, the form of the Yukawa matrices as defined in the `Substitutions` section might not be preserved along the RG-flow. In other words, some components of the Yukawa matrices which were set to `0` in the model file may have a non-vanishing $\beta$-function starting from the 1- or 2-loop order. In such cases, PyR@TE will generate a warning message in the Latex output, informing the user that the RG-flow is inconsistent.

(4) The last possibility consists in defining new quantities, expressed in terms of the couplings of the model. For instance, if the user wishes to obtain the $\beta$-functions in terms of the $\alpha_i = g_i^2/(4\pi)$ instead of the coupling constants $g_i$, he/she would have to define the following substitutions:

```
Substitutions: {
    alpha1 : g1**2 / (4 pi),
    alpha2 : g2**2 / (4 pi),
    alpha3 : g3**2 / (4 pi)
}
```

In this situation, PyR@TE will internally take care of computing the $\beta$-functions of the new couplings, using the relation

$$\beta(\alpha_i) = \frac{1}{4\pi} 2 g_i \beta(g_i),  \tag{33}$$

and then substituting $g_i^2$ with $4\pi\,\alpha_i$ everywhere in the expression of the $\beta$-functions.

**Latex** – In this section, the user is invited to define any substitution that should be performed in the Latex output. Although it is possible to choose as the name of a coupling a simple Latex expression (*e.g.* `g_1` ), we recommend to avoid it in practice (and more generally to avoid the use of any special character in the name of the couplings) since it might lead to some unexpected behavior. Instead, the mapping to Latex expressions should be defined in the dedicated `Latex` section, so the user has a full control on the output.

*3.3. Running PyR@TE*

As for previous versions of the software, PyR@TE 3 can be run either from the console or from an interactive IPython notebook. In the former case, the general syntax is:

`$ python pyR@TE.py -m [path to model file] [-args]`

We emphasize that the `python` alias must be associated with a Python 3 executable correctly linked to the dependencies listed in section 3.1. In an interactive IPython notebook, PyR@TE would be run using:

`%run pyR@TE.py -m [path to model file] [-args]`

The only mandatory argument is the `-m` argument, used to specify the path (relative or absolute) to the model file. The list of optional arguments is presented in Table 1. Compared to the previous versions, the number of available command-line arguments was reduced in order to simplify to a maximum the use of the software. Instead, most of the settings are gathered in a configuration file, `'default.settings'`, which can be found in PyR@TE's main directory. The list of options that are available in this file are listed in Table 2. For some of the options, there is an overlap between the default and command-line settings. In this case, the latter always have the priority over the former. For example, even if the gauge invariance of the model is to be checked by default, the user may use the `-no-gi` command-line option to prevent this behavior.

We note in passing that the gauge invariance check of the model is a new feature in PyR@TE 3. When enabled, the invariance of all types of coupling under infinitesimal gauge transformations is tested. For instance, in the scalar quartic sector, gauge invariance requires that

$$(T_\phi^A)_{ae}\lambda_{ebcd} + (T_\phi^A)_{be}\lambda_{aecd} + (T_\phi^A)_{ce}\lambda_{abed} + (T_\phi^A)_{de}\lambda_{abce} = 0 \tag{34}$$

for all $a, b, c, d$ spanning the space of the scalar components of the model. It is strongly advised to check the gauge invariance of a model at least once, after which the option `-no-gi` may be used to speed up the computation[3].

The `-l` and `DefaultLoopLevel` settings should be followed either by a positive integer (1, 2 or 3) or the keyword `max`, meaning that the computation should be carried out at the maximum loop-order, *i.e.* 3-loop for the gauge couplings and 2-loop for the other couplings. Another possibility is to use a list of three elements[4], *e.g.* `[3,2,1]` to set the respective orders of the gauge, Yukawa, and quartic couplings RGEs to different values.

The `-rb` and `RealBasis` settings enable the use of a new feature available in PyR@TE 3, namely the rotation of the generators of real representations. These representations have the property that there exists a set of bases where their generators $t^a$ are imaginary, skew-symmetric matrices. In the following, such bases will be referred to as *real bases*. In addition, for adjoint representations (which are always real), there exists a basis where the usual relation

$$(t^a)^i{}_j = -i f^{ai}{}_j, \tag{35}$$

with $f^{ai}{}_j$ the structure constants of the Lie algebra, is satisfied. However the generators of real representations originally computed by PyLie never satisfy the two above properties, preventing the user to properly implement models where one or several fields transform under real representations. In order to circumvent this issue, a new algorithm was implemented in PyLie, allowing to systematically rotate the generators of real representations to a real basis; and in the case of adjoint representations, to the basis where Eq. (35) is satisfied. To trigger this behavior, the `-rb` (or `RealBasis`) option may be assigned three possible values:

- '`-rb adjoint`' rotates only adjoint representations,
- '`-rb all`' rotates all real representations,
- '`-rb False`' or '`-rb None`' disables the rotation to real bases.

We emphasize that complex and pseudo-real representations are unaffected by this setting.

The `EndCommands` default settings allows to run one or several shell commands after the run. It was designed to compile the Latex output automatically after its generation by PyR@TE. The general syntax is:

    EndCommands "command 1, command 2, ..."

For instance, we may call `pdflatex` twice using

    EndCommands "pdflatex [name].tex, pdflatex [name].tex"

where `[name]` will be automatically replaced by the name of the model (and therefore of the output files) before running the commands.

---

[3] For large gauge groups and/or representations, the gauge invariance check may take up to a few minutes due to the large number of tensor contractions to be computed. For simpler models, the computation takes a few seconds at most.

[4] On some operating systems, the use of square brackets in the command line may not work properly. In this case, putting the list inside quotation marks should solve the problem.

| Option | Short-hand | Description |
| --- | --- | --- |
| `--Loops` | `-l` | Set the loop-level of the computation |
| `--Results` | `-res` | Set the directory in which the output is stored |
| `--RealBasis` | `-rb` | Set the behavior regarding the generators of real representations |
| `--Quiet` | `-q` | Switch off informative output |
| `--no-KinMix` | `-no-kin` | Neglect the Abelian kinetic mixing |
| `--CheckGaugeInvariance` | `-gi` | Switch on the Lagrangian gauge invariance check |
| `--no-CheckGaugeInvariance` | `-no-gi` | Switch off the Lagrangian gauge invariance check |
| `--LatexOutput` | `-tex` | Switch on the Latex output |
| `--no-LatexOutput` | `-no-tex` | Switch off the Latex output |
| `--MathematicaOutput` | `-math` | Switch on the Mathematica output |
| `--no-MathematicaOutput` | `-no-math` | Switch off the Mathematica output |
| `--PythonOuput` | `-py` | Switch on the Python output |
| `--no-PythonOuput` | `-no-py` | Switch off the Python output |

Table 1: Command-line options in PyR@TE 3. The second part of the table shows the various switches which may be used by the user to override some of the default settings.

| Option | Description |
| --- | --- |
| `VerboseLevel` | Set the verbose level : `Info`, `Critical` (only errors and warnings). |
| `ResultsFolder` | Set the default folder in which the results are stored. |
| `LogFolder` | Set the default folder in which the log files are stored. |
| `DisableLogFiles` | `True/False`. |
| `DefaultLoopLevel` | Default loop level of the computation. |
| `CheckGaugeInvariance` | `True/False`. |
| `PrintComputationTimes` | Display the computation times : `True/False`. |
| `RealBasis` | Default behavior regarding the generators of real representations. |
| `CreateFolder` | Store the output in a folder named after the model: `True/False`. |
| `CopyModelFile` | Copy the original model file in the results folder : `True/False`. |
| `LatexOutput` | `True/False`. |
| `AbsReplacements` | In the Latex output, replace $x^*x$ with $|x|^2$: `True/False`. |
| `GroupTheoryInfo` | In the Latex output, display some information about the gauge groups and their representations: `True/False`. |
| `MoreGroupTheoryInfo` | Display information about the first `N` representations of the gauge groups: `N/False`. |
| `MathematicaOutput` | `True/False`. |
| `MathematicaSolver` | In the Mathematica output, generate a ready-to-use RGE solver: `True/False`. |
| `PythonOutput` | `True/False`. |
| `EndCommands` | Commands to be automatically run from the console after the output is generated. |

Table 2: Default settings available in PyR@TE's configuration file.

*3.4. Output*

After the computation, PyR@TE may generate three types of output, namely Latex, Mathematica, and Python files. With the `CreateFolder` option enabled, all the output files will be gathered in a folder named after the model (*e.g.* SM/ in our case).

The Latex output contains a detailed summary of the content of the model (gauge groups, particles, Lagrangian, substitutions), and the results of the computations, *i.e.* the $\beta$-functions of the various couplings of the model. If the `GroupTheoryInfo` setting is enabled, some information about the gauge groups and their representations will be added in an appendix, along with the expressions of the CGCs possibly used in the definition of the Lagrangian. Using the `MoreGroupTheoryInfo` setting followed by a positive integer will display information about the first few representations of the gauge groups, in addition to the representations actually populated by the fields of the model.

The Mathematica output consists in a single file, containing the expressions of the $\beta$-functions. In addition, if the `MathematicaSolver` default setting is set to `True`, the file is enhanced with a ready-to-use RGE solver. To solve the RGEs, the first thing that the user needs to do is to define the boundary conditions, that is the value of the couplings of the model at some initial energy scale. Then, after having defined the running scheme (the loop level for each type of couplings), the user may solve the system of RGEs using the `RGsolve[]` function and finally plot the results using the `RGplot[]` function.

Finally, the Python output consists of three files, gathered in a `PythonOutput` folder. For instance, in the case of the SM:

- `RGEs.py` contains the expressions of the $\beta$-functions in the form of Python functions.

- `SM.py` contains various classes used to define the couplings of the model, the RGEs, and several functions used to solve the RGEs and plot the running couplings. In principle, this file should not be modified by the user, but may still serve as a basis for a more sophisticated analysis.

- `run.py` is the file that should be modified and run by the user to perform the RG analysis. For more details about the use of the classes and functions called in this file, we invite the reader to refer to the documentation provided with PyR@TE in the `doc/` folder.

## 4. Validation

The output of PyR@TE 3 was validated against some of the results found in the literature, in particular in the 3-loop gauge sector which is a new feature compared to the previous version. Full agreement was found between PyR@TE 3 and the following references :

- SM gauge couplings RGEs at three loop [25, 26, 27].

- Gauge, Yukawa and quartic sector of the general Two-Higgs-doublet model at respective orders 3-2-1 [28, 29].

- 2-loop dimensionless and dimensionful RGEs of the toy model described in [14] in the SUSY limit[5].

- 1-loop quartic RGEs of simple $SU(5)$ and $SO(10)$ models [30].

- 1-loop Yukawa sector of the minimal $SU(5)$ model [31].

- 1-loop quartic, trilinear and scalar mass RGEs of the most general Georgi-Machacek model [32]. We find agreement for the 1-loop gauge coupling RGEs after correcting a trivial mistake in Eq. (C1) of [32].

In addition, we compared the results of PyR@TE 3 against those of PyR@TE 2 for some of the models presented in [10] and [11]. Doing so, we were able to compare the performance of the two versions of the software in terms of execution time, as illustrated in Table 3. We first present the comparison of the results for some of the models listed in [10]:

- Standard Model with a real scalar singlet : full agreement at 2-loop.

- Standard Model with a complex scalar doublet : full agreement at 2-loop.

- Standard Model with Majorana triplet fermion and Dirac doublet : full agreement in the gauge, Yukawa and quartic sectors at 2-loop. Agreement is found for the scalar mass RGEs after correcting a mistake in the latest version of the PyR@TE 2 code. For the fermion mass RGEs, we find agreement once the corrected formulae presented in [14] are taken into account.

- SM extended by a complex triplet and vectorlike doublets : at 1-loop, we find many disagreements in the quartic sector due to an incomplete implementation of the model. Indeed, in PyR@TE 1 it was not possible to contract the same set of fields in different ways. As a consequence, couplings (operators) allowed by gauge invariance were missing. In such a situation, the resulting RGEs are arbitary and vary depending on the choice of the $n$ legs chosen to compute the various $n$-point contributions to the $\beta$-functions (see also footnote [5]).

The limitation of PyR@TE 1 mentioned in the last point was overcome in PyR@TE 2, as illustrated in [11] with the example of a toy model where the SM is extended with a scalar triplet. For this model, the scalar potential is given by Eq. (30), and we were able to compare the results given by PyR@TE 2 and 3. Full agreement is found at 2-loop, after an obvious error in the model file from PyR@TE 2 is corrected (an additional lepton doublet was unintentionally included).

Finally, we validated the output of PyR@TE 3 for the $U(1)_{B-L}$ extension of the SM described in [11]. In the gauge, Yukawa and quartic sector, full agreement is found between the two versions of the code in presence of kinetic mixing in the Abelian sector. However, there is a disagreement for the scalar mass $\beta$-functions starting at 2-loop which we describe in the following. Letting

$$V \supset \mu_H H^\dagger H + \mu_\chi \chi^\dagger \chi \tag{36}$$

be the quadratic part of the scalar potential $V$, where $H$ is the usual SM Higgs doublet and $\chi$ is a complex scalar field which is a SM singlet with (B-L)-charge 2, the disagreement occurs in the two-loop contributions of $\beta(\mu_H)$ and $\beta(\mu_\chi)$. More precisely, the problematic terms are of the generic form $g^4 \mu_H$ and $g^4 \mu_\chi$, respectively, so we only include the terms of these forms in the expressions below. We note that the gauge coupling matrix is taken to be

$$\begin{pmatrix} g_1 & g_{12} \\ g_{21} & g_{BL} \end{pmatrix}, \tag{37}$$

while $g_2$ stands for the $SU(2)_L$ coupling constant.

---

[5]As explained in [14], the toy model in its non-SUSY form is not well defined, leading to arbitrary results depending on details of the implementation. It is possible to get full agreement between the results from SARAH and PyR@TE before taking the SUSY limit by a simple mapping. However, we refrain from providing these details here since they concern unphysical results and the mapping is only simple because the implementation conventions used in SARAH and PyR@TE are quite similar.

| Model | Loop order | PyR@TE 2 | PyR@TE 3 |
|---|---|---|---|
| SM B-L | 1 | 114 | 1.5 |
|  | 2 | 8823 | 11 |
|  | 2 + 3 (gauge) | / | 23 |
| SM + complex triplet | 1 | 385 | 1.0 |
|  | 2 | 59936 | 3.2 |
|  | 2 + 3 (gauge) | / | 5.7 |
| SM + scalar singlet | 1 | 79 | 0.9 |
|  | 2 | 5765 | 4.3 |
|  | 2 + 3 (gauge) | / | 5.6 |
| SM + complex doublet | 1 | 153 | 1.2 |
|  | 2 | 39666 | 6.2 |
|  | 2 + 3 (gauge) | / | 9.4 |
| SM + Majorana triplet + Vectorlike doublet | 1 | 262 | 1.3 |
|  | 2 | 15653 | 10.7 |
|  | 2 + 3 (gauge) | / | 13.2 |

Table 3: Comparison of the running times of PyR@TE 2 and PyR@TE 3 for several models, using a machine with a 1.8 GHz Intel Core i7 processor. The running times are expressed in seconds.

**Expressions of $\beta^{(2)}(\mu_H)$**

Using PyR@TE 2:

$$\frac{1}{\mu_H} \beta^{(2)}(\mu_H) \supset \frac{1105}{96} g_1^4 - \frac{145}{16} g_2^4 + \frac{1105}{96} g_{12}^4 + \frac{7}{2} g_1^2 g_2^2 + \frac{1105}{48} g_1^2 g_{12}^2 + 17 g_1^2 g_{21}^2 + 17 g_{12}^2 g_{BL}^2 \tag{38}$$
$$+ \frac{7}{2} g_{12}^2 g_2^2 + \frac{40}{3} g_1^3 g_{21} + \frac{40}{3} g_{12}^3 g_{BL} + \frac{40}{3} g_1^2 g_{12} g_{BL} + \frac{40}{3} g_1 g_{12}^2 g_{21} + 34 g_1 g_{12} g_{21} g_{BL}$$

Using PyR@TE 3:

$$\frac{1}{\mu_H}\beta^{(2)}(\mu_H) \supset \frac{557}{48}g_1^4 - \frac{145}{16}g_2^4 + \frac{557}{48}g_{12}^4 + \frac{15}{8}g_1^2 g_2^2 + \frac{557}{24}g_1^2 g_{12}^2 + 17g_1^2 g_{21}^2 + 17g_{12}^2 g_{BL}^2 \quad (39)$$
$$+ \frac{15}{8}g_{12}^2 g_2^2 + \frac{40}{3}g_1^3 g_{21} + \frac{40}{3}g_{12}^3 g_{BL} + \frac{40}{3}g_1^2 g_{12} g_{BL} + \frac{40}{3}g_1 g_{12}^2 g_{21} + 34 g_1 g_{12} g_{21} g_{BL}$$

**Expressions of $\beta^{(2)}(\mu_\chi)$**

Using PyR@TE 2:

$$\frac{1}{\mu_\chi}\beta^{(2)}(\mu_\chi) \supset 648 g_{BL}^4 + 648 g_{21}^4 + \frac{422}{3}g_1^2 g_{21}^2 + \frac{422}{3}g_{12}^2 g_{BL}^2 + 1296 g_{21}^2 g_{BL}^2 + \frac{640}{3}g_1 g_{21}^3$$
$$+ \frac{640}{3}g_{12} g_{BL}^3 + \frac{640}{3}g_1 g_{21} g_{BL}^2 + \frac{640}{3}g_{12} g_{21}^2 g_{BL} + \frac{844}{3}g_1 g_{12} g_{21} g_{BL} \quad (40)$$

Using PyR@TE 3:

$$\frac{1}{\mu_\chi}\beta^{(2)}(\mu_\chi) \supset 672 g_{BL}^4 + 672 g_{21}^4 + \frac{422}{3}g_1^2 g_{21}^2 + \frac{422}{3}g_{12}^2 g_{BL}^2 + 1344 g_{21}^2 g_{BL}^2 + \frac{640}{3}g_1 g_{21}^3$$
$$+ \frac{640}{3}g_{12} g_{BL}^3 + \frac{640}{3}g_1 g_{21} g_{BL}^2 + \frac{640}{3}g_{12} g_{21}^2 g_{BL} + \frac{844}{3}g_1 g_{12} g_{21} g_{BL} \quad (41)$$

Although the above expressions alone do not allow one to discard any of the two sets of results, there is a fact indicating that PyR@TE 2 is internally inconsistent when it comes to kinetic mixing: running PyR@TE 2 with the option `--KinMix` to disable the effects of kinetic mixing, or taking the limit $g_{12}, g_{21} \to 0$ in the two expressions above do not yield the same result, as would be expected. Interestingly, the results obtained with the `--KinMix` option (*i.e.* disabling kinetic mixing in PyR@TE 2) completely agree with those of PyR@TE 3 in the limit where $g_{12}$ and $g_{21}$ vanish.

As a concluding remark, while we cannot make the claim that the results obtained with PyR@TE 3 are correct[6], the treatment of kinetic mixing for the scalar mass RGEs in PyR@TE 2 appears to be inconsistent. Furthermore, the implementation of kinetic mixing in the formalism described in Sec. 2 is quite natural whereas it has been incorporated a posteriori in PyR@TE 2 [11] by enhancing [13] the expressions derived by Machacek & Vaughn[7]. As a result, the implementation of the kinetic mixing in PyR@TE 2 is quite involved and we were not able to directly identify the source of the inconsistency. In any case, the process of validation performed above allowed us to highlight at least two flaws present in the PyR@TE 2 code, and we recommend to systematically use PyR@TE 3 in the future instead of its previous versions which are now deprecated.

## 5. Conclusion and outlook

In this paper, we have presented a new version of PyR@TE, a Python tool for the computation of renormalization group equations for general, non-supersymmetric gauge theories. Compared to the previous versions PyR@TE 1 [10] and PyR@TE 2 [11], an important number of improvements have been realized. First of all, the new PyR@TE 3 is now compatible with Python 3. This is an important step since Python 2 has reached "End Of Life status" as of January 2020 and it is officially recommended to switch to Python 3 now. Furthermore, large parts of the code have been rewritten. Its new core relies on a new formalism developed by Poole & Thomsen [1] to compute the $\beta$-functions for all dimensionless couplings. The corresponding expressions for the dimensionful couplings have recently been derived in [22] completing the set of expressions in this formalism. In this framework, gauge kinetic mixing is naturally implemented, and the Weyl consistency relations between gauge, quartic and Yukawa couplings [17, 19, 20] are automatically satisfied. One of the main new features is the possibility for the user to compute the gauge coupling $\beta$-functions up to the three-loop order. Therefore, it is now possible with PyR@TE 3 to study the RGEs at (3, 2, 1) loop order for the gauge, Yukawa, and quartic couplings, respectively, as has been advocated in the literature [34]. Much effort was put towards the development of a faster code. As a consequence, there is a *drastic* improvement in performance as can be inferred from Tab. 3. This is of great practical importance. For one-loop calculations PyR@TE 3 is roughly a factor 100 faster than PyR@TE 2 and for the more complex two-loop calculations PyR@TE 3 is 1000 to 10000 times faster than PyR@TE 2. In all examples presented in Tab. 3, PyR@TE 3 just needs a couple of seconds to compute the RGEs – even in the case of three-loop gauge couplings. For more involved models, in particular ones with widely extended scalar sectors, the typical running times are at most of a few minutes at the maximum loop-orders.

---

[6]It is worth noting that complete agreement was found with Thomsen's code RGBeta [33], based on the same formalism.

[7]A similar comment holds for the Mathematica package SARAH [9].

There is a number of other improvements which are noteworthy: (i) We have worked to make the model file more flexible and user-friendly (albeit largely but not entirely backward compatible). (ii) The PyLie module has been modified not only to improve its speed performance but also to include some new features. For example, PyR@TE 3 can now handle all simple Lie algebras including the symplectic and exceptional ones. There is also now the option to rotate the basis in the target space of the representation such that the generators for a real representation are imaginary and skew-symmetric. (iii) The two-loop RGEs for VeVs [23, 24] have been implemented in PyR@TE 3. (iv) We also note in passing that an interface between PyR@TE 3 and the Mathematica package Feynrules [35] is under development which will be described elsewhere.

We have validated PyR@TE 3 against the literature for several models (see Sec. 4) and found complete agreement. We have also compared the results of PyR@TE 3 with the ones of PyR@TE 2 for some of the models already discussed in [10, 11]. Again, we have found largely agreement up to two problems which could be associated with the PyR@TE 2 code. For several reasons we recommend to use PyR@TE 3 in the future: PyR@TE 2 (and Python 2) are no longer developed/maintained, the new formalism in the PyR@TE 3 code has a very clear structure with a natural implementation of kinetic mixing, there are several new features which make PyR@TE 3 more user-friendly and efficient.

## Acknowledgements

We are indebted to Florian Lyonnet from whom we have taken over the development of PyR@TE. This work would not have been possible without the foundations laid by him in the previous versions of the code. We are also grateful to Colin Poole and Anders Eller Thomsen for many useful discussions. Finally, we would like to thank Jan Kwapisz, Aaron Held and Fabien Besnard for their helpful contributions in testing and validating PyR@TE 3. We also wish to thank Tom Steudtner for his help with the validation of the SUSY toy model from [14]. This work was supported in part by the IN2P3 project "Théorie – BSMGA".

## Appendix A. Kinetic mixing in the Abelian sector

We consider in this appendix a theory with $p$ Abelian gauge factors. Prior to the redefinition of the gauge fields by absorption of the gauge couplings, the Abelian mixing is parametrized by the mixing matrix $\xi$:

$$\mathcal{L}_{\text{kin}} \supset -\frac{1}{4} F_{\mu\nu}^{\text{T}} \xi F^{\mu\nu} \,. \tag{A.1}$$

$\xi$ is a $p \times p$ symmetric matrix satisfying $\xi_{mm} = 1$. Therefore it contains $p(p-1)/2$ degrees of freedom. Initially, the gauge coupling matrix $g$ in the Abelian sector is diagonal:

$$g = \begin{pmatrix} g_1 & & \\ & \ddots & \\ & & g_p \end{pmatrix} \,. \tag{A.2}$$

Following the approach presented in [13, 11], we may perform a redefinition the gauge fields $V' = \xi^{1/2} V$ such that $\xi$ does no longer explicitly appear in the expression (A.1). Accordingly, we define an extended gauge coupling matrix $G_{\text{mix}}$:

$$G_{\text{mix}} \equiv g \xi^{-1/2} \,. \tag{A.3}$$

Going one step further, we may absorb the gauge couplings by redefining the gauge fields in the following way:

$$V'' = G_{\text{mix}} V' \,, \tag{A.4}$$

so that equation (A.1) can eventually be put in the form

$$\mathcal{L}_{\text{kin}} \supset -\frac{1}{4} F_{\mu\nu}^{\text{T}} H^{-2} F^{\mu\nu} \,, \tag{A.5}$$

where

$$H^2 = G_{\text{mix}} G_{\text{mix}}^{\text{T}} \,. \tag{A.6}$$

Clearly, according to the above definition, $H^2$ is invariant under rotations $G_{\text{mix}} \to G_{\text{mix}} O$ where $O$ is a $p \times p$ orthogonal matrix. Such a rotation enables one to transform $G_{\text{mix}}$ into an upper-triangular matrix where its $p(p+1)/2$ degrees of freedom ($p$ gauge couplings plus $p(p-1)$ mixing parameters) appear explicitly [11].

## Appendix B. The SM model file

As an illustration of the overall structure of a model file in PyR@TE 3, we show below the full SM model file. We note that this file can be found in PyR@TE's `models/` subdirectory, along with several BSM model files.

```yaml
# YAML 1.1
---
Author: Lohan Sartore
Date: 08.06.2020
Name: SM
Groups: {U1Y: U1, SU2L: SU2, SU3c: SU3}

Fermions: {
    Q : {Gen: 3, Qnb: {U1Y: 1/6, SU2L: 2, SU3c: 3}},
    L : {Gen: 3, Qnb: {U1Y: -1/2, SU2L: 2}},
    uR : {Gen: 3, Qnb: {U1Y: 2/3, SU3c: 3}},
    dR : {Gen: 3, Qnb: {U1Y: -1/3, SU3c: 3}},
    eR : {Gen: 3, Qnb: {U1Y: -1}},
}

RealScalars: {
}

ComplexScalars: {
    H : {RealFields: [Pi, Sigma], Norm: 1/sqrt(2), Qnb: {U1Y: 1/2, SU2L: 2}},
}

Potential: {

    Definitions: {
        Htilde[i] : Eps[i,j]*Hbar[j]
    },

    Yukawas: {
        Yu : Qbar[i,a] Htilde[i] uR[a],
        Yd : Qbar[i,a] H[i] dR[a],
        Ye : Lbar[i] H[i] eR
    },

    QuarticTerms: {
        lambda : (Hbar[i] H[i])**2
    },

    ScalarMasses: {
        mu : -Hbar[i] H[i]
    }

}

Vevs: {
    v: Pi[2]
}

Substitutions: {
    # Rename the gauge coupling constants
    g_U1Y : g1,
    g_SU2L : g2,
    g_SU3c : g3,

    # Possibly define GUT normalizations
```

```
        g1 : sqrt(5/3) * g1
}

Latex: {
    # Particles

    uR : u_R,
    dR : d_R,
    eR : e_R,

    Pi : \Pi,
    Sigma : \Sigma,

    Htilde : \tilde{H},

    # Couplings

    g1 : g_1,
    g2 : g_2,
    g3 : g_3,

    Yu : Y_u,
    Yd : Y_d,
    Ye : Y_e,

    lambda : \lambda,
    mu : \mu
}
```

### Appendix C. Group theory and PyLie

This appendix is dedicated to the description of the group theoretical functionalities available in PyR@TE 3. All the group theory related computations are handled, since the previous version of PyR@TE, by the PyLie module. PyLie is essentially a Python rewrite of the group theory module of SUSYNO [36]. Such computations comprise for instance the possibility of computing the generators of a given representation, the structure constants of the Lie algebra, or the Clebsch-Gordan coefficients (CGCs) of a given set of representations. A lot of effort was put in the development of PyR@TE 3 to improve the performances of PyLie's main functions. This concerns in particular the three kinds of calculations listed above. However, despite these efforts, some calculations may be quite time-consuming when it comes to high-dimensional representations. Therefore, based on the previous version of PyLie, we developed a database in which the results of the group-theoretical computations are systematically stored for any later use. At various steps of the computation of the RGEs, PyR@TE 3 interacts with this database through the `PyLieDB` module. The user may interact as well with this database through an interactive IPython session or a Jupyter notebook in order to access the results of PyLie's computations.

After a discussion about the Dynkin labels as a way to uniquely identify the representations of a given gauge group, we introduce the use of CGCs to build a Lagrangian in PyR@TE 3. A short tutorial (in the form of an interactive Python notebook) available in the `doc/` repository of PyR@TE 3 is dedicated to the interaction with PyLie's database. In order to keep this appendix as concise as possible, we invite the interested user to refer to this tutorial if needed.

*Appendix C.1. Dynkin labels*

The Dynkin labels of an irreducible representation are a set of $N$ positive integers, where $N$ is the rank of the algebra. They characterize the decomposition of the highest weight of the representation in terms of the $N$ fundamental weights of the algebra. In PyLie, they are used to identify uniquely the representations of a given Lie algebra. In practice, the Dynkin labels take the form of Python lists, and can be used in the model file in place of the usual notation for representations based on their dimensions. For instance, one could define the quantum numbers of the quark doublet `Q` as:

```
Q: {Gen: 3, Qnb: {U1Y: 1/6, SU2L: [1], SU3c: [1,0]}}
```

We invite the user who is not familiar with the Dynkin labels notation to refer to the document provided with PyR@TE 3 in the `doc/` directory, where the first few representations of some usual gauge groups are listed along with their Dynkin labels. Several functions were also implemented in the `PyLieDB` module that can be called by the user to get information on the representations and their Dynkin labels.

*Appendix C.2. CGCs*

In some cases, it may not be possible to express a given gauge invariant combination of the fields using a notation with contracted indices involving only the gauge generators, the Levi-Civita tensor and the fields themselves. In such cases, the user may use the Clebsch-Gordan coefficients (CGCs) generated by PyLie.

Given a set of $N$ fields $F_{k,\,k\leq N}$ transforming under the irreducible representations $r_{k,\,k\leq N}$ of a given Lie algebra, we call CGCs all the tensors $I^A_{i_1,\ldots,i_N}$ such that the quantity

$$I^A_{i_1,\ldots i_N} F_1^{i_1} \ldots F_N^{i_N} \tag{C.1}$$

is gauge invariant. Since in PyR@TE we consider only renormalizable theories, $N$ may only equal 2, 3 or 4. We note that there are $M$ linearly independent CGCs, where $M$ is the multiplicity of the trivial representation in the decomposition of the tensor product $r_1 \otimes \ldots \otimes r_N$.

One of the main functionalities of PyLie consists in finding a basis of CGCs given a set of $N$ fields. These CGCs may be used by the user in the model file to build the Lagrangian. To this end, the CGCs must be defined in the `Definitions` section of the model file using the two following possible syntaxes :

```
Definitions: {
    C : cgc(groupName, field_1, ... field_N, P),
  # or, equivalently,
    C : cgc(group, representation 1, ... representation N, P)
}
```

The last argument, `P`, is an integer indicating that we are asking for the Pth CGCs returned by PyLie. If omitted, the first invariant tensor will be returned. Defined this way, `C` is a tensor with `N` indices that may be used in the expression of the Lagrangian as any other tensor quantity defined in the model file.

In order to illustrate the use of CGCs in a simple case, let us first consider the example of the up-type Yukawa coupling in the SM. In a notation with contracted indices, one would simply write:

```
Yukawas: {
    Yu : Qbar[i,a] Htilde[i] uR[a]
}
```

where `Htilde` is the conjugated Higgs field, defined in the `Definitions` section of the SM model file. The $SU(2)$ and $SU(3)$ contractions rely on the two simple decompositions

$$SU(2) \quad : \quad \mathbf{2} \otimes \mathbf{2} = \mathbf{3} \oplus \mathbf{1}, \tag{C.2}$$

$$SU(3) \quad : \quad \mathbf{\bar{3}} \otimes \mathbf{3} = \mathbf{8} \oplus \mathbf{1}, \tag{C.3}$$

from which it can be seen that only one gauge singlet may be constructed in each case. Using the CGCs instead, the `Yu` term would be defined as:

```
Definitions: {
# SU2 contraction
    c1 : cgc(SU2L, Qbar, Hbar),
  # or, equivalently,
    c1 : cgc(SU2, -2, -2),

# SU3 contraction
    c2 : cgc(SU3c, Qbar, uR),
  # or, equivalently,
    c2 : cgc(SU3, -3, 3)
},

Yukawas: {
    Yu : c1[i,j] c2[a,b] Qbar[i,a] Hbar[j] uR[b]
}
```

Of course, the above example looks like an unnecessary complication since `c1` and `c2` actually correspond to the rank 2 Levi-Civita tensor and the 3-dimensional Kronecker delta.

As a more sophisticated example, let us now consider a $SU(5)$ model with the following field content:

$$\begin{aligned} \text{Fermions} \quad &: \quad \psi_5 \sim \mathbf{\bar{5}} \quad \text{and} \quad \psi_{10} \sim \mathbf{10} \\ \text{Scalars} \quad &: \quad \phi \sim \mathbf{5} \end{aligned} \tag{C.4}$$

In order to construct the Yukawa sector, we make use of the decompositions

$$\mathbf{\bar{5}} \otimes \mathbf{\bar{5}} \otimes \mathbf{10} = \mathbf{126} \oplus \mathbf{75} \oplus \mathbf{24} \oplus \mathbf{24} \oplus \mathbf{1}, \tag{C.5}$$

$$\mathbf{10} \otimes \mathbf{5} \otimes \mathbf{10} = \mathbf{\overline{175'}} \oplus \mathbf{\overline{126}} \oplus \mathbf{75} \oplus \mathbf{75} \oplus \mathbf{24} \oplus \mathbf{24} \oplus \mathbf{1}, \tag{C.6}$$

enabling one to define the two following Yukawa couplings

$$Y_5\,\psi_5\phi^\dagger\psi_{10} \quad \text{and} \quad Y_{10}\,\psi_{10}\phi\psi_{10}\,. \tag{C.7}$$

The two above terms are usually expressed in a simple form, noticing that the 10-plet can be reorganized in a $5\times 5$ antisymmetric matrix. However, in PyR@TE, the field $\psi_{10}$ only carries a single $SU(5)$ index and we must use the CGC notation in order to build the Yukawa Lagrangian. We show below a minimal implementation of this $SU(5)$ toy model.

```
Groups: {SU5: SU5}

Fermions: {
    psi5  : {Gen: 3, Qnb: {SU5: -5}},
    psi10 : {Gen: 3, Qnb: {SU5: 10}}
}

ComplexScalars: {
    phi : {RealFields: [phiR, phiI], Norm: 1/sqrt(2), Qnb: {SU5: 5}}
}

Potential: {
    Definitions: {
        c5  : cgc(SU5, psi5, phibar, psi10),
        c10 : cgc(SU5, psi10, phi, psi10)
    },

    Yukawas: {
        Y5  : c5[i,j,k] psi5[i] phibar[j] psi10[k],
        Y10 : {c10[i,j,k] psi10[i] phi[j] psi10[k], symmetric}
    }
}
```